\documentclass[fleqn,12pt]{article}
\usepackage{amssymb,amsfonts,amsthm,amsmath,graphicx,url}
\usepackage[longnamesfirst]{natbib}
\usepackage{longtable,booktabs}
\usepackage[dvipsnames]{xcolor}
\usepackage[normalem]{ulem}
\usepackage{setspace}
    \doublespace
\usepackage{lineno}
\usepackage{xr}
\RequirePackage[vmarginratio=1:2, outer=35mm, inner=25mm, lines=25]{geometry}

\usepackage{fullpage}

\RequirePackage{lipsum}
\RequirePackage{appendix}

\newcommand{\beq}{\begin{eqnarray*}}
\newcommand{\eeq}{\end{eqnarray*}}
\newcommand{\beqq}{\begin{equation}}
\newcommand{\eeqq}{\end{equation}}

\newcommand{\whpsi}{\widehat{\psi}}
\newcommand{\whp}{\widehat{p}}

\newcommand{\wtpsi}{\widetilde{\psi}}

\newcommand*\patchAmsMathEnvironmentForLineno[1]{
  \expandafter\let\csname old#1\expandafter\endcsname\csname #1\endcsname
  \expandafter\let\csname oldend#1\expandafter\endcsname\csname end#1\endcsname
  \renewenvironment{#1}
     {\linenomath\csname old#1\endcsname}
     {\csname oldend#1\endcsname\endlinenomath}}
\newcommand*\patchBothAmsMathEnvironmentsForLineno[1]{
  \patchAmsMathEnvironmentForLineno{#1}
  \patchAmsMathEnvironmentForLineno{#1*}}
\AtBeginDocument{
\patchBothAmsMathEnvironmentsForLineno{equation}
\patchBothAmsMathEnvironmentsForLineno{align}
\patchBothAmsMathEnvironmentsForLineno{flalign}
\patchBothAmsMathEnvironmentsForLineno{alignat}
\patchBothAmsMathEnvironmentsForLineno{gather}
\patchBothAmsMathEnvironmentsForLineno{multline}
}

    \title{{\bf Two-stage approaches to the analysis of occupancy data I: The homogeneous case} \\
    (Analysis of occupancy data)
    }    
\author{{\bf N.\ Karavarsamis\footnote{Corresponding author: N. Karavarsamis, \emph{email}: nkarav$@$unimelb.edu.au}  \hspace*{0.2mm} and R.\ M.\ Huggins}\\
School of Mathematics and Statistics \\
The University of Melbourne, Victoria 3010, Australia. \\
}
\date{}

\begin{document}
   
\maketitle    
\thispagestyle{empty} 
    
\noindent A\textsc{bstract}: \hspace{2mm}
         Occupancy models are used in statistical ecology to estimate species dispersion.
         The two components of an occupancy model are the detection and occupancy
         probabilities, with the main interest being in the occupancy
         probabilities.
         We show that for the homogeneous occupancy model there is an orthogonal transformation of the parameters that gives a natural  two-stage inference procedure based on a conditional likelihood. We then extend this to a partial likelihood that gives explicit estimators of the model parameters. By allowing the separate modelling of the detection and occupancy probabilities, the extension of the two-stage approach to more general models
         has the potential to simplify the computational routines used there.
    
    \vspace*{5mm}
    \noindent K\textsc{ey} W\textsc{ords}: \hspace{2mm} Imperfect detection; Occupancy models; Orthogonal parameterisation; Partial likelihood.
    
\clearpage    
    \section{Introduction}

     The importance of occupancy studies in conservation planning, biodiversity monitoring, and invasive species biology, is well known \citep{hoffman10} and a complete
    description of hierarchical occupancy models is contained in \citet{royle08}. 
     These data consist of observations on whether a site is occupied and are collected over repeated visits to a number of sites.
    They require less
    survey effort and are cheaper than studies that estimate abundance \citep{pollock02,mac02, wintle05}. The full
    likelihood \citep{mac02} involves both the probability a site is occupied and the probability of detection of a species at a site if it is occupied.

As the full likelihood involves both the detection and occupancy probabilities, in general inference requires the joint modelling of these quantities and 
    can quickly become quite complicated. For example if there are 10 covariates then there are $2^{20}=1048576$ possible joint models for detection and occupancy.
    However, if the modelling can be done separately for each of detection and occupancy there are $2^{10}+2^{10}= 2048$ models. 
    Even with a smaller number of covariates, if one fits non-parametric functions of the covariates rather than simple linear models the number of parameters
    to be estimated can become large.
    Moreover, computer algorithms to 
    fit models with large numbers of parameters can be unstable and finding global rather than local maxima can be difficult.
    As a first step towards simplifying the likelihood with the aim of eventually alleviating these
computational difficulties we consider the simple homogeneous model for both occupancy and detection probabilities.

In statistics there are several ways of simplifying complex likelihoods. One is through parameter orthogonalization \citep[e.g.][]{Cox:1987wo}.
 We consider a simple transformation of the parameters that yields two orthogonal parameters. We show that the resulting estimate of the detection probability arises from a conditional likelihood and that of the constructed parameter arises from a simple binomial likelihood, giving a natural two stage procedure to compute the maximum likelihood estimates.  
Partial likelihood is also used to simplify
complex likelihoods and it's theory is well developed \citep[e.g.][]{cox75,wong86, gill92}. 
We consider a partial likelihood approach that yields the equivalent of two binomial likelihoods and simple analytic estimates of both parameters.

In Section \ref{sec-note} we give our notation and the  likelihood of \citet{mac02}. In Section \ref{sec-orthog} we give the orthogonal transformation and  show this yields a conditional likelihood
to estimate the detection probabilities.
In Section \ref{sec-partial} we give a partial likelihood approach that further simplifies estimation. In Section \ref{sec-sims}  we conduct some simulations to examine the efficiency of the resulting occupancy estimator.
In Section \ref{sec-applications} we apply the estimators to several data sets. Some technical results are given in the Appendices.
    
    \section{Notation and Likelihood}\label{sec-note}
    
    Consider $S$ sites labelled $s=1,\dots,S$ where each site is visited on $\tau$ occasions.
    We suppose the occupancy status of each site is constant over the visits.
    Let $\psi$ be the probability a site  is occupied and 
    $p$ be the probability the species is observed at a site on a given occasion
    given it is present. 
    Then $\theta=1-(1-p)^\tau$ is the probability of at least one
    detection at a site  given the site is occupied.
    Let $y_s$ denote the number of occasions upon which the species was
    detected at site $s$ and let $y=\sum_{s=1}^S y_s$ be the total number of detections, let $f_0$ be the number of sites where none of the species was detected
    and let $O=S-f_0$ be the number of sites where they were.
    Reorder the $S$ sites $s=1,\dots,O, O +1, \ldots S$, where $1, \ldots, O$ denote the sites at which at least one detection occurred and $O +1, \ldots S$ the remaining sites at which no sightings occurred.
    The full likelihood \citep{mac06} is then
    \begin{align}
    L(\psi,p)&\propto (1-\psi\theta)^{f_0}\psi ^{S-f_0}\prod_{s=1}^O {\tau \choose y_s} p^{y_s}(1-p)^{\tau-y_s}\nonumber\\
    &\propto (1-\psi\theta)^{f_0}\psi ^{S-f_0} p^y (1-p)^{O\tau-y}.\label{eq-lik0}
    \end{align}
    This likelihood may be maximised numerically and the maximum likelihood estimates $\widehat\psi$ and $\widehat p$ found, for example using
    the \texttt{R} package \texttt{unmarked}  \citep{fiske14}.

    \section{Orthogonal Transformation and Conditional Likelihood}\label{sec-orthog}

  For the homogeneous occupancy model there is a natural transformation that yields orthogonal parameters.
    Let $\eta=\psi\theta$ be the probability that occupancy is detected at a site and now consider the parameters $\eta,p$.
    The full likelihood (\ref{eq-lik0}) is then proportional to
    \begin{eqnarray}
    L(\eta,p)
    &=&(1-\eta)^{f_0}\eta^{S-f_0}\frac{\prod_{s=1}^O  p^{y_s}(1-p)^{\tau-y_s}}{\theta^{S-f_0}}.\nonumber\\
    &=&(1-\eta)^{f_0}\eta^{S-f_0} \frac{p^y(1-p)^{O\tau-y}}{\theta^O}\label{eq-orthog}
   \end{eqnarray}
   To determine orthogonality of $\eta$ and $p$ we need to examine the resulting information matrix \citep{Cox:1987wo}.
   The log-likelihood is
    \begin{align}
    \ell(\eta,p)&=f_0 \log(1-\eta) + (S-f_0)\log(\eta)\label{eq-eta}\\
    &+ y \log(p)+(O\tau-y)\log(1-p)
     -O\log(\theta).\label{eq-p}
    \end{align}
    Hence
   \begin{align}
    \frac{\partial \ell(\eta,p)}{\partial \eta}&=-\frac{f_0}{(1-\eta)}+\frac{(S-f_0)}{\eta},\label{eq-score-eta}\\
   \frac {\partial \ell(\eta,p)}{\partial p}&=\frac{y}{p} -\frac{(O\tau-y)}{(1-p)}
    -\frac{O\tau(1-\theta)}{(1-p)\theta}\label{eq-score-p}
    \end{align}
     so that
   $ 
{\partial^2 \ell(\eta,p)}/{\partial \eta \partial p}= {\partial^2 \ell(\eta,p)}/{\partial p \partial \eta}=0
   $,
    and the transformed parameters $\eta$ and $p$ are orthogonal.

    From (\ref{eq-score-eta}) it is seen that the mle of $\eta$ is
    $
    \widehat\eta={(S-f_0)}/{S}=O/S
    $. 
  The maximum likelihood estimator of $p$ maximises (\ref{eq-p}), which is the log-likelihood conditional on at least one detection at each site.
  It can be computed using the {\tt VGAM} package in {\tt R} \citep{yee10}.
Let $\widehat p$ be the resulting estimator of $p$ and let $\widehat \theta=1-(1-\widehat p)^\tau$.  The invariance property of maximum likelihood estimates yields that the mle of $\psi$ is $\widehat\psi=\widehat\eta/\widehat\theta$. As these are the maximum likelihood estimates the usual estimates of their variances may be used.
Thus for the homogeneous model maximum likelihood estimation for the occupancy model may be naturally conducted as a two-stage procedure.

{\bf Remark} A direct verification of the equivalence of the maximum likelihood and conditional likelihood estimates is possible. Note that
$
\log L(\psi,p)=f_0\log(1-\psi\theta)+(S-f_0)\log(\psi)+y \log(p)+(O\tau-y)\log(1-p)
$ and
\begin{align}
     \frac{\partial L(\psi,p)}{\partial \psi}&= \frac{S(1-\psi\theta)-f_0}{\psi(1-\psi\theta)}\label{eq-1}\\
  \frac{\partial L(\psi,p)}{\partial p}&=  \frac{y}{p}-\frac{O\tau-y}{1-p}-\frac{f_0\psi\tau(1-\theta)}{(1-p)(1-\psi\theta)}\nonumber\\
  &=  \frac{y}{p}-\frac{O\tau-y}{1-p}-\frac{f_0\psi\theta}{(1-\psi\theta)O}\frac{O\tau(1-\theta)}{(1-p)\theta}.\label{eq-2}
\end{align}
Now, setting (\ref{eq-1}) equal to 0 and solving for $\psi$ yields
$
\psi={O}/{S\theta}
$
and substituting this into 
$
{f_0\psi\theta}/\{(1-\psi\theta)O\}
$
yields
$
{f_0O\theta}/\{\theta S(1-O/S)O\}={f_0}/{(S-O)}=\frac{f_0}{f_0}=1
$.   Thus, setting (\ref{eq-1})  and (\ref{eq-2}) to zero are equivalent to setting (\ref{eq-score-eta})
and (\ref{eq-score-p}) to zero.

 \section{Partial Likelihood}\label{sec-partial}

We have seen the conditional likelihood can be fitted in the {\tt VGAM} package in {\tt R}. However, this package is not yet as sophisticated as the
{\tt glm}  function in {\tt R}.
Hence, to further simplify estimation we exploit that for occupancy models there are repeated observations at each site so that there is more information on the detection
probabilities than on the occupancy probabilities. 
Let  $b_s$
be the number of occasions remaining after the first detection at site $s$ and let $b=\sum_{s=1}^S b_s$. The number of re-detections at site $s$ is just  $y_s-1$  so  that the total number of re-detections is $y - O$.
Let $a=O\tau-O-b$ be the total over the sites where occupancy was detected of the number of occasions before the first detection.
Then $(\ref{eq-orthog})$ may be written as
\begin{align}
   L(\eta,p)&= (1-\eta)^{f_0}\eta^{S-f_0}\label{eq-L1}\\
   &\times \frac{p^O(1-p)^a}{\theta^{O}}\label{eq-junk}\\
   &\times p^{y-O} (1-p)^{b-(y-O)}.\label{eq-L2}
\end{align}
In this decomposition (\ref{eq-L1}) and (\ref{eq-L2}) are proportional to simple binomial likelihoods
and we base inference on these two components. The component (\ref{eq-junk}) is proportional to the 
product of the probabilities of an individuals first detection time given they are detected at least once.
In our partial likelihood approach we ignore this component. That is, this partial likelihood approach  ignores the information 
 up to and including  the first occasion at which a sighting was made at each site.

Using the partial likelihood (\ref{eq-L2}) to estimate $p$ yields that the partial likelihood estimator of $p$ is
$
\tilde p = (y-O)/{b}$ and the usual Binomial variance ${\rm Var} (\tilde p) = \tilde p (1- \tilde p) /b$. 
To estimate $\psi$ we then estimate $\eta$ from (\ref{eq-L1}) as in Section \ref{sec-orthog} then back transform to yield
$\widetilde \psi = \widehat \eta / \widetilde \theta =(S-f_0)/S\widetilde \theta$, where $\widetilde \theta=1-(1-\tilde p)^\tau$.
In Appendix \ref{app-var} we show that
\begin{equation}
{\rm Var}\left(\widetilde \psi  \right)\approx
\left(\frac{\psi(1-\psi\theta)}{S\theta}+\psi^2\right) \frac{
  \tau^2(1-p)^{2(\tau-1)}}{\theta^2} \frac{p(1-p)}{b}+\frac{\psi(1-\psi\theta)}{S\theta}.\label{eq-varpsi-hom}
\end{equation}

	\begin{table}[ht]
	\centering
	\begin{tabular}{|l|rr|rr||rr|rr|}
	  \hline
	&\multicolumn{2}{c|}{Partial}&\multicolumn{2}{c||}{Full}&\multicolumn{2}{c|}{Partial}&\multicolumn{2}{c|}{Full}\\
	  \hline
	\rule{0pt}{3ex} & $\whp$ & $\whpsi$ &$\whp$ & $\whpsi$ & $\whp$ & $\whpsi$ &$\whp$ & $\whpsi$ \\
	 \hline
	$S=1000$,\ $\tau=5$ & 0.100 & 0.400 & 0.100 & 0.400 & 0.050 & 0.400 & 0.050 & 0.400 \\ \hline
	Median estimate & 0.100 & 0.401 & 0.100 & 0.402 & 0.049 & 0.407 & 0.049 & 0.407 \\
	Median SE & 0.016 & 0.058 & 0.015 & 0.057 & 0.016 & 0.124 & 0.015 & 0.120 \\
	MAD & 0.015 & 0.060 & 0.015 & 0.057 & 0.016 & 0.127 & 0.015 & 0.123 \\
	Efficiency & 0.915 & 0.925 &  & & 1.021 & 0.988 &  &   \\
	   \hline
	$S=100$,\ $\tau=5$ & 0.200 & 0.400 & 0.200 & 0.400 & 0.200 & 0.600 & 0.200 & 0.600 \\ \hline
	 Median estimate & 0.199 & 0.405 & 0.197 & 0.411 & 0.198 & 0.609 & 0.197 & 0.609 \\
	  Median SE & 0.049 & 0.091 & 0.045 & 0.088 & 0.040 & 0.106 & 0.037 & 0.101 \\
	MAD & 0.051 & 0.093 & 0.045 & 0.089 & 0.039 & 0.103 & 0.036 & 0.101 \\
	  Efficiency & 0.801 & 0.914 &  & & 0.843 & 0.909 &  & \\ \hline
	\end{tabular}
	\caption{Simulation results to compare our partial likelihood and full
	  likelihood approaches. We report the  median of the estimated values, the median of the estimated standard errors, the median absolute deviation of the estimates and efficiency  computed from the robust measures as ratio of variances.}
	\label{tab-eff}
	\end{table}

\section{Simulations}\label{sec-sims}

As the conditional likelihood approach to estimating the detection probabilities that arose from the orthogonal transformation yields the mle's there is
no need to evaluate it in simulations.
To examine the efficiency of the partial likelihood  approach, we
simulate an experiment and compute the full and  partial likelihood
estimates. The standard errors of the full maximum likelihood
estimates were computed using a numerically computed observed
information matrix. 
We conducted 1000 simulations for each parameter combination and considered $S=1000$ and 100, $\psi=0.4$ and 0.6, and $p=0.05, 0.1$ and 0.2.
In  Table \ref{tab-eff} we report the medians of the estimates, the associated
standard errors and the  median absolute deviation\footnote{The MAD is corrected by {\tt R} for asymptotically normal consistency by a factor of 1.4826.} (MAD) of the 
partial and full likelihood estimators for $\tau=5$. We report the median, rather than the mean, so our results were not unduly influenced by the occasional outlying estimate.
However, in line with convention the efficiency  was computed using the variances of the estimates.
There is no indication of bias in either method, both
estimated standard errors appear reliable and the efficiency of our method for $\psi$ is above 90\%.  
Further, simulations to examine the effects of small probabilities, and different numbers of
    occasions, and the performance of the partial likelihood estimator, in a setting similar to that in our example, are in Appendix \ref{app-res}. In the first instance, there was evidence of some bias
    for small values of $p$ and small $\tau$
    and the estimated standard errors appear slightly too large. In  the latter setting,
    the partial likelihood estimator performs well, the estimated standard errors appear quite reasonable and the estimator of $\psi$, is quite efficient. 

\section{Application}\label{sec-applications}

The application consists of  detections of the Growling Grass Frog (\emph{Litoria raniformis}) that were collected at $S=27$ sites with  $\tau=4$ occasions during  the 2002--2003 season, as part of a larger study \citep{heard06}. For these data, $f_0=12$ and $y= 47$.
Here the method of Section \ref{sec-partial}
yielded $\tilde p = 0.889$ with estimated standard error $0.052$ and
$\widetilde \psi=0.556$ with estimated standard error $0.096$. The full
likelihood, with these estimates used as starting
values,  yielded $\whp = 0.780$ with estimated standard error $0.054$ and
$\whpsi=0.557$ with estimated standard error $0.096$. 
The method of Section \ref{sec-orthog} yielded the same values.
The
difference between the two estimates of the detection probability may
be of interest. One interpretation is that there is some difference between the
initial detection probability and that on subsequent occasions, as the
full likelihood includes information from the first captures whereas
the partial likelihood does not.

\section{Discussion}\label{sec-disc}

Two stage procedures are not new in statistical inference.
For example  in other contexts \citet{Sanathanan:1972kl,Sanathanan:1977cv}  developed  two-stage procedures that she termed conditional likelihood and  \citet{Huggins:1989hb} has employed a two-stage procedure to estimate
population size using capture-recapture data where the first stage employed a conditional likelihood to estimate the capture
probabilities.
We have focused on  estimating the occupancy and detection probabilities in the simple homogeneous case. Firstly, after using an orthogonal transformation we have a natural
two-stage method to compute the maximum likelihood estimates. 
This also provides an interesting example where the maximum likelihood estimator of the detection probability $p$ is also
a conditional maximum likelihood estimator. This may also be useful to illustrate parameter orthogonality and conditional and partial likelihood approaches to inference.
The orthogonality of $\eta$ and $p$ implies that: the mle's $\widehat \eta$ and $\widehat p$ are asymptotically independent, the asymptotic standard error of $\widehat\eta$
is the same whether $p$ is known or unknown, there may be simplifications in the numerical derivation of the estimates and the mle $\widehat\eta(p)$  of $\psi$ when $p$ is given varies slowly as a function of $p$ \citep{Cox:1987wo}.
In our case, the last three points are emphasised as $\widehat \eta$ does not depend on $p$.
The resulting sensitivity of $\widehat \psi$ to changes in $\widehat p$ is examined in Appendix \ref{app-sens}.
Both two-stage procedures offer some simplifications. The conditional likelihood approach allows separate estimation of the detection probability
and the occupancy probability. However, it requires a relatively non-standard conditional likelihood to estimate the detection probabilities.
This can be implemented in the {\tt R} package {\tt VGAM}. 
The two-stage partial likelihood approach further simplifies the derivation of the estimates giving explicit  estimators of both the  occupancy and detection probabilities.
The cost is a small loss of efficiency.
The difference between the partial and conditional likelihood approaches is in how the detection probability $p$ is estimated.
We have restricted ourselves to the homogeneous case, but the extension to more complex models where occupancy and detection probabilities both depend on covariates has potential to reduce the computational burden and increase the computational efficiency of maximum likelihood estimation there. This will be examined elsewhere.

\section*{Acknowledgements}

We would like to thank Dr Geoffrey Heard (School of Botany, The University of Melbourne) for supplying us with the data for the Growling Grass Frog.

   \bibliography{Sep2017}
    \bibliographystyle{apalike}

    \clearpage

    \appendix
    
    \section{Simulation Results}\label{app-res}
    
    To examine the effects of small probabilities and different numbers of
    occasions in our method, we took $S=100$, $p=0.05$, $\psi=0.6$ and considered both $\tau=5$ and $\tau=10$.
    The results are summarized in Table \ref{tab-small}  where we only
    report simulations with $\widehat \psi < 1$. 
    Next, we took $S=27$, $\tau=4$, $\psi=0.6$ and $p=0.6$, which is
    similar to the values in the Growling Grass Frog application.
    The results are reported in Table \ref{tab-aseg} along with those for the full likelihood.

\begin{table}[h!]
\centering
\begin{tabular}{|l|rr|rr|}
  \hline
&\multicolumn{2}{c}{$\tau=5$}&\multicolumn{2}{|c|}{$\tau=10$}\\\hline
 &$p$ & $\psi$ &$p$ & $\psi$\\
  \hline
True value & 0.050 & 0.600 & 0.050 & 0.600 \\\hline
Median estimate & 0.067 & 0.475 & 0.052 & 0.587 \\
Median SE & 0.046 & 0.326 & 0.020 & 0.208 \\
MAD & 0.031 & 0.227 & 0.017 & 0.182 \\
   \hline
\end{tabular}
\caption{Simulations in the homogeneous case for the two-stage estimator with a moderate number ($S=100$) of sites, small ($\tau=4$) and large ($\tau=10$) 
numbers of occasions and
  small values of $p$. We give the true values of the parameters, the median of the estimated values, the median of the estimated standard errors and the median absolute deviation (MAD) of the estimates.}
\label{tab-small}
\end{table}

  	\begin{table}[h!]
  	\centering
  	\begin{tabular}{|l|rr|rr|}
  	  \hline
  	&\multicolumn{2}{c}{Partial}&\multicolumn{2}{|c|}{Full}\\ \hline
  	 &$p$ & $\psi$ &$p$ & $\psi$\\
  	  \hline
  	True Value & 0.600 & 0.600 & 0.600 & 0.600 \\ \hline
  	Median estimate & 0.600 & 0.604 & 0.600 & 0.604 \\
  	Median SE & 0.078 & 0.097 & 0.066 & 0.097 \\
  	MAD & 0.078 & 0.104 & 0.065 & 0.105 \\\hline
  	Efficiency & 0.709 & 0.991 &  &  \\
  	   \hline
  	\end{tabular}
  	\caption{Comparisons of the two-stage (partial likelihood) and the full maximum likelihood estimates in the homogeneous case for a small number of sites ($S=27$), small
  	  number of occasions ($\tau=4$) and large values
  	  of $p$. We give the true values of the parameters, the median of the estimated values, the median of the estimated standard errors and the median absolute deviation (MAD) of the estimates.}
  	\label{tab-aseg}
  	\end{table}

    \section{Variance of $\widetilde \psi$}\label{app-var}
    
    Firstly the variance of $\hat p$ is relatively straightforward.
     Given $b$, $y-O \sim {\rm Bin}(b,p)$ so
    	  that
    	$E(\tilde p | b)=p$ and ${\rm Var}(\tilde p | b)=p(1-p)/b$. Hence
    	$
    	{\rm Var}(\tilde p)=p(1-p)E(1/b)
    	$, which we estimate by $S^2_p=\tilde p
    (1-\tilde p)/b$.
    To determine the variance of $\widetilde \psi$ let $\overline \psi= (S-f_0)/(S \theta)$ be the mle for known $p$. Now, $f_0 \sim {\rm Bin}(S,1-\psi\theta)$ so that  $\overline\psi$ is unbiased
    	and has variance $V=\psi(1-\psi\theta)/(S\theta)$. 
    To find the variance of  $\widetilde \psi$, write
    	$
    	\widetilde \psi = \overline \psi \theta/\widetilde\theta$ and
        a Taylor expansion yields
    	$\widetilde\theta\approx \theta+\tau(1-p)^{\tau-1}(\tilde p-p)$. Then
    	$
    	\wtpsi \approx \overline \psi \left\{1-{\tau(1-p)^{\tau-1}(\tilde p-p)}/{\theta}\right\}.
    	$
    	Thus, 
    	$
    	E\left(\widetilde \psi \big| b,f_0 \right)\approx \overline \psi
    	$,
    	$
    	E\left( \overline\psi^2 \right)=\psi(1-\psi\theta)/(S\theta)+\psi^2
    	$, so that
    	$
    	{\rm Var}\left(\widetilde \psi \big| b,f_0\right)\approx {\overline\psi^2
    	  \tau^2(1-p)^{2(\tau-1)}}{\theta^{-2}} {\rm Var}\left(\tilde p \big| b,n_0\right)
    	$.
      This yields (\ref{eq-varpsi-hom}).
      
      \begin{figure}[h!]
      \begin{center}
      \includegraphics[scale=0.6]{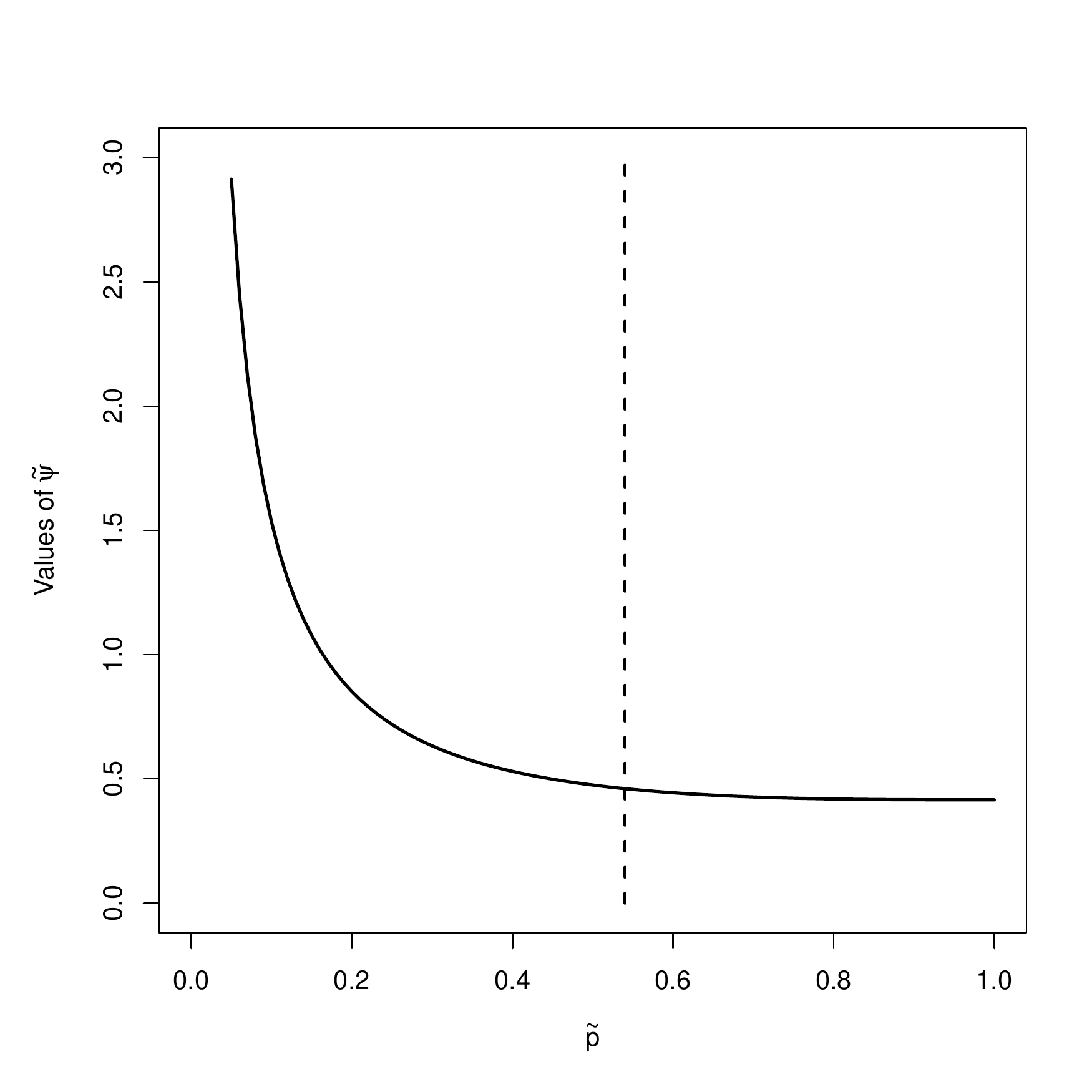}
      \end{center}
      \caption{Values of $\whpsi_p$ plotted against $p$ from the partial likelihood score equations for an example with $S = 77 $ and $\tau= 3$. The vertical dashed line indicates $\whp = 0.54$. The horizontal dashed line shows minimum value for $\whpsi_p$, at 0.415. }
      \label{psi-pplot}
      \end{figure}
    
    \section{Sensitivity}\label{app-sens}
    
    Small differences in the estimated detection probabilities had little effect on the estimated occupancies. 
    Let $\overline\psi_p$ be the estimated value of $\psi$ for given $p$. Then, $\partial \overline \psi_p / \partial p=(S-f_0)(1-p)^{(\tau-1)}\tau/((S(1-(1-p)^\tau))$ which is usually small and when multiplied by the difference in the estimated probabilities is even smaller still. 
    To illustrate this suppose $S= 77$ and $\tau = 3$, $f_0 = 45$ and $y=57$. This gave $\tilde p = 0.54$. 
    In Figure
     \ref{psi-pplot} we give  a plot of $\overline \psi_p$ against or $p\in (0,1)$.
    It is clear that when $p$ is small (i.e.\ $p \leq 0.15$) values for $\overline \psi_p$ are greater than 1. For values of $p \ge 0.3$ there is little change in $\overline \psi_p$ and for $p \ge 0.5$ practically no change in $\whpsi_p$. Hence, when estimates for $p$, $\whp$, are obtained from our method that differ to estimates given by the full likelihood, there is little to no change in  $\overline \psi_p$.

\end{document}